\documentclass{icrc}

\usepackage{times}
\usepackage{graphicx} 

\begin{document}

\title{Test of primary model predictions by EAS size spectra}
\author[1]{S.V. Ter-Antonyan}
\affil[1]{Yerevan Physics Institute, 2 Alikhanian
Br. Str., 375036 Yerevan, Armenia}
\author[2]{P.L. Biermann}
\affil[2]{Max-Planck-Institute f\"ur Radioastronomie
Auf dem H\"ugel 69, D-53121 Bonn, Germany}

\correspondence{samvel@jerewan1.yerphi.am}

\firstpage{1}
\pubyear{2001}


\maketitle

\begin{abstract}
High statistical accuracy of experiments KASCADE and ANI allowed
to obtain approximations of primary energy spectra and elemental
composition in the "knee" region. Obtained results point out to
the correctness of QGSJET interaction model and 2-component model
of primary cosmic ray origin up to $100$ PeV energies.
\end{abstract}

\section{Introduction}
Absolute differential EAS size spectra around the knee measured at
different atmosphere depths and different zenith angles are not
explained yet from the point of view of a single $A-A_{Air}$ 
interaction model and a single model of primary energy spectra
and elemental composition. Such an attempt has been made in 
work \citep{TH}
based on an unified analysis of KASCADE, AKENO, EAS-TOP and ANI
EAS size
spectra. The results of approximations of primary energy spectra 
by rigidity-dependent steepening spectra  pointed to the 
correctness of QGSJET interaction model and two-component composition of
primary proton spectrum in the knee region.\\
Here, on the basis of KASCADE \citep{KAS} and ANI 
 \citep{ANI} EAS size spectra the multi-component
model of primary cosmic ray origin \citep{PB} has been tested in
the framework of method \citep{TH}.   

\section{Method}
The testing of primary energy spectra was carried out in
$0.03-500$
PeV primary energy range using $\chi^2$-minimization \citep{TH}
\begin{equation}
\min\{\chi^{2}/ \xi\} \equiv
\min\Big\{\frac{1}{\xi}
\sum_{i}^{m}\sum_{j}^{n}
\frac{(f_{i,j}-F_{i,j})^{2}}{\sigma_{f}^{2}+\sigma_{F}^{2}}
\Big\}
\end{equation} 
where
\begin{equation}
f_{i,j}\equiv\frac { \partial I(E_{e},\overline{\theta_j},t) } 
{\partial N_{e,i}^* }
\end{equation}
is detectable EAS size spectra measured in $i=1,\dots m$ size
intervals and $j=1,\dots n$ zenith angular intervals 
(see Fig.1, symbols),
$E_{e}$ is an energy threshold of detected EAS electrons,
$N_{e,i}^*(E>E_{e})$ is the estimation value of EAS sizes
obtained by detected electron lateral distribution functions at
observation level ($t$);\\
\begin{equation}
F_{i,j}\equiv  
\sum_{A}
\int_{E_{min}}^{\infty}
\frac{\partial \Im_{A}} {\partial E_{0}}
W_{\theta} (E_{0},A,N_{e}^{*},\overline\theta ,t)
dE_{0}  
\end{equation}
are the expected EAS size spectra at $\partial \Im_{A}/ \partial E_{0}$
energy spectra of primary nuclei ($A=1,\dots,59$);\\
$\xi=m\cdot n -p-1$ is a degree of freedom at $p$ number of
unknown parameters.\\
The function $W_{\theta}$ in expression (3) is determined in general
case as
\begin{equation}
W_{\theta}\equiv
\int_{\theta _1}^{\theta _2}
\int_{0}^{\infty}
\frac{\partial\Omega(E_{0},A,\theta ,t)}{\partial N_{e}}
\frac{P_{\theta}}{\Delta_{\theta}}
\frac{\partial \Psi(N_{e})}{\partial N_{e}^{*}}
\sin\theta d\theta dN_{e}
\end{equation}
where $\partial\Omega / \partial N_{e}$  is an EAS size spectrum at 
the observation level $(t)$ for given $E_{0},A,\theta$ parameters 
of a primary nucleus and depends on $A-A_{Air}$ interaction model;\\
$\Delta_{\theta}=\cos\theta_{1}-\cos\theta_{2}$;
\begin{equation}
P_{\theta}\equiv
\frac{1}{X\cdot Y}
\int\!\!\int D(N_{e},E_{0},A,\theta,x,y)dxdy
\end{equation}
is a probability to detect an EAS by scintillation array
at EAS core coordinates $|x|<X/2$, $|y|<Y/2$ and to obtain estimations
of  
EAS parameters ($N_{e}^{*}$, $s$ - shower age, $x^{*},y^{*}$ - shower core 
location)
with given accuracies; \\
$\partial \Psi / \partial N_{e}^{*}$ is a
distribution 
of $N_{e}^{*}(N_{e},s,x,y)$ for given EAS size ($N_{e}$).\\

In most of EAS experiments the EAS cores are selected in $P\simeq1$
range providing a log-Gaussian form of a measuring
error  ($\partial\Psi /\partial N_{e}^{*}$) with an average value
$\ln(N_{e}\cdot\delta)$ and RMSD $\sigma_{N}$,
where $\delta$ involves all transfer factors (an energy threshold
of detected EAS electrons, $\gamma$ and $\mu$ contributions) and
slightly depends on $E_{0}$ and $A$. In these cases, one may 
standardize the measured EAS size spectra to the EAS size spectra at the
observation level
\begin{equation}
\frac{\partial I(0,\overline{\theta},t)} {\partial N_{e}}\simeq
\eta\frac{\partial I(E_{e},\overline{\theta},t)} {\partial N_{e}^{*}},
\end{equation}
where
$\eta=\delta^{(\gamma_{e}-1)}\exp\{(\gamma_{e}-1)^{2}\sigma_{N}^{2}/2\}$
and  $\gamma_{e}$ is the EAS size power index.\\                           
Taking the above into account, the 
expected EAS size spectra (3) can be estimated according to
the expression
\begin{equation}
F_{i,j} =
\eta\sum_{k=1}^{k_{\max}}
\int_{E_{min}}^{\infty}
\frac{\partial \Im _{A_{k}}}{\partial E_{0}}
\frac{\partial\Omega(E_{0},\overline{A}_{k},\overline{\theta},t)}{\partial
N_{e}}dE_{0}            
\end{equation}
where the sum is performed into limited number ($k_{\max}$) of
nuclear group.

\subsection{2-component primary energy spectra}
Energy spectra of primary nuclei ($A\equiv1-59$)  
according to the multi-component model of primary cosmic ray
origin 
\citep{PB} are presented in 2-component form:  
\begin{equation}
\frac{\partial \Im} {\partial E_{A}}= 
\Phi_{A}\Big(
\delta_{A,1}\frac{d\Im_{1}}{dE_{A}}+
\delta_{A,2}\frac{d\Im_{2}}{dE_{A}}
\Big) 
\end{equation} 
where the first component (ISM) is
derived from the explosions of normal supernova into an
interstellar medium with expected rigidity-dependent power law
spectra \citep{PB} 
\begin{equation} 
\frac{d\Im_{1}}{dE_{A}} = \left\{
\begin{array}{l@{\quad:\quad}l} 
E_{A}^{-\gamma_{1}} &
E_{A} < E_{ISM} \\
0 & E_A>E_{ISM} 
\end{array} \right.
\end{equation} 
and the second component (SW) is a result of the explosions of
stars into their former stellar winds with expected
rigidity-dependent power law spectra \citep{PB} 
\begin{equation}
\frac{d\Im_{2}}{dE_{A}} = 
\left\{ \begin{array}{l@{\quad:\quad}l}
E_{A}^{-\gamma_{2}} & E_{A} < E_{SW} \\
E_{SW}^{-\gamma_{2}}(E_{A}/E_{SW})^{-\gamma_{3}} &
E_A>E_{SW} \end{array} \right. 
\end{equation} 
where $\Phi_{A}$ is a scale factor ($E$ in TeV units) from
approximations \citep{BWPB};
\begin{equation} 
E_{ISM}=R_{ISM}\cdot Z, \;\;\; E_{SW}=R_{SW}\cdot Z
\end{equation} 
are the corresponding rigidity-dependent cut-off energies of
ISM-component and knee energies of SW-component;\\
$R_{ISM}$ and $R_{SW}$ are model parameters of magnetic rigidities 
of corresponding components and $Z$ is the charge of $A$ nucleus.\\
The values of model predictions \citep{PB} for spectral
parameters are:\\
$\gamma_1=2.75\pm0.04$, $\gamma_2=2.67\pm0.03$, $\gamma_3=3.07\pm0.1$ \\
and rigidities $R_{ISM}\simeq120$ TV, $R_{SW}\simeq 700$ TV
at factors of uncertainty $\sim 2$.\\
The fractions of each component
($\delta_{A,i}\equiv\delta_{i}(E_{A},A)$, $i=1,2$) are determined according
to 
\begin{equation}
\delta_{A,1}=1-\delta_{A,2} 
\end{equation}
\begin{equation}
\delta_{A,2}=(2ZR_{ISM})^{(\gamma_{2}-\gamma_{A,0})} 
\end{equation}
at $\gamma_{A=1,0}=2.75$ and $\gamma_{A>1,0}=2.66$ \citep{PB}.
The expressions (12,13) are consequences of normalization of 
(8-10) to approximation of balloon and satellite data
\citep{BWPB} at $E_{A}=1$ TeV.\\

 \begin{figure}[t]
 \includegraphics[width=8.5cm,height=12cm]{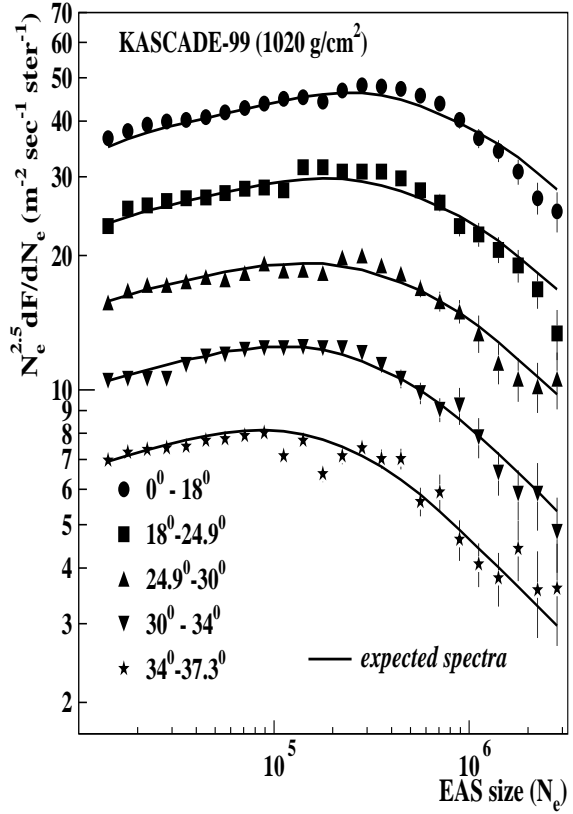} %
 \caption{KASCADE EAS size spectra at different zenith angles 
\citep{KAS} (symbols). Lines correspond to expected EAS size
spectra
according to QGSJET interaction model and 2-component origin of 
primary cosmic rays.}
 \end{figure}
Thus, minimizing $\chi^2$-functional (1) on the basis of measured 
values of $\partial I(\overline{\theta}_{i})/ \partial N_{e,j}$
and corresponding expected EAS size spectra (7) at given $m$ zenith 
angular intervals and $n$ EAS size intervals one may evaluate  
parameters of primary spectra for given ($k_{\max}$) nuclear
groups. 
Evidently, the accuracies of solutions for spectral
parameters strongly depend on the number of measured intervals 
($m\cdot n$), statistical errors and correctness of
$\partial\Omega(E_{0},A,\theta,t)/\partial N_{e}$ 
determination in the framework of a given interaction model. Moreover, 
the value of $\chi^{2}$ points out a reliability of applying
primary model.\\
Differential EAS size spectra $\partial\Omega (E_{0},A,\theta ,t)/
\partial N_{e}$ for given
$E_{0}\equiv 0.032,0.1,\dots,100$ PeV,
$A\equiv1$,$4$,$12$,$16$,$28$,$56$, 
$t\equiv 0.5,0.6,\dots,1$ Kg/cm$^{2}$,
$\cos\theta\equiv 0.8,0.9,1$
were calculated using CORSIKA562(NKG) EAS simulation code (Heck et
al.,1998) at
QGSJET \citep{QGS} interaction model.
Intermediate values are calculated using
4-dimensional log-linear interpolations. Estimations of errors of
expected EAS size spectra $\partial\Omega /\partial N_{e}$
at fixed $E_{0},A,\theta,t$ parameters did not exceed $3-5\%$.\\
 \begin{figure}[t]
 \includegraphics[width=8.3cm,height=10cm]{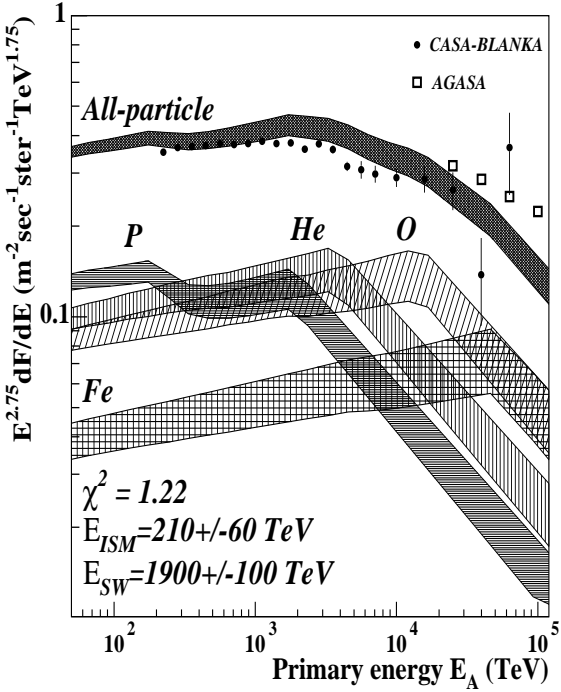} %
 \caption{Expected primary all particle and $P,He,O,Fe$ energy spectra
obtained by approximation of KASCADE data.
The symbols are CASA-BLANCA \citep{CB} and AGASA \citep{AG} data.}
 \end{figure}
Unknown parameters in minimization (1) were:\\
$E_{ISM}$ - cut-off energy of ISM proton component;\\
$E_{SW}$ - knee energy of proton SW-component;\\
$\gamma_{3}$ - power index after the knee ($E_{SW}$);\\ 
$\eta$ - a systematic shift (discrepancy) from expression (6).\\
Moreover, power indices $\gamma_{1,2}$ were changed too in the
range of the model uncertainty and the relative uncertainties of 
expected spectra ($\sigma_F$) in the $\chi^2$-minimization (1)
we set equal to 3\%.

\section{Results}
 \begin{figure}[t]
 \includegraphics[width=8.3cm,height=9cm]{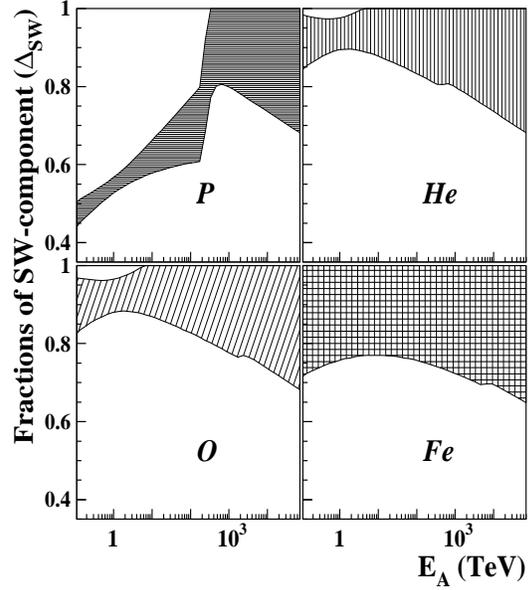} %
 \caption{Expected fractions of SW-component at different nuclei
obtained by approximation of KASCADE EAS size spectra.}
 \end{figure}
The testing of primary model predictions by minimizations (1) 
were carried out on the basis of KASCADE \citep{KAS} ($t=$1020
g/cm$^{2}$) and ANI \citep{ANI} (700 g/cm$^{2}$) EAS size
spectra at 5 zenith angular intervals.
The KASCADE EAS size spectra (symbols) and corresponding expected spectra
by 2-component model (lines) are shown in Fig.1. 
The obtained
primary all-particle energy spectrum and energy spectra of 
$P$, $He$, $O$, $Fe$ nuclear groups are presented in Fig.2 in
comparison with CASA-BLANCA \citep{CB} and AGASA \citep{AG} measurements. 
The obtained 
values of approximation parameters of primary energy spectra are:\\
$\gamma_1=2.78\pm0.03$; $\gamma_2=2.65\pm0.03$;
$\gamma_3=3.28\pm.07$;\\
$E_{ISM}=210\pm60$ TeV; $E_{SW}=1900\pm100$ TeV;\\
and $\eta=1.03\pm0.03$ at $\chi^2=1.22$.\\
Obtained fractions of the primary stellar wind (SW) component
\begin{equation}
\Delta_{SW}(E_A,A)=\delta_{A,2}\frac{\partial\Im_{2}/\partial E_A}
{\partial \Im/\partial E_A}
\end{equation}
versus energy at different primary nuclei are presented in Fig. 3. The results are  
extrapolated up to $0.1$ TeV primary energy range.\\
The testing of 2-component model of CR origin was also carried out
by ANI EAS size spectra \citep{ANI} measured at the mountain
level.
The results are shown in Fig. 4 and correspond to $\chi^2=1.05$ and
systematic shift $\eta=1.14$. The systematic underestimations
of KASCADE ($\eta=1.03$) and ANI ($\eta=1.14$) EAS size spectra
(see expression (6)) one may explain by energy thresholds of
detected EAS electrons ($E_e\sim3$ MeV for KASCADE and 
$E_e\sim10$ MeV for ANI experiments).\\
The values of spectral parameters obtained by approximations of 
ANI data agree with corresponding parameters obtained from
approximations of KASCADE data except for cut-off
energy $E_{ISM}=460\pm100$ TeV at ANI data analysis.

\balance
\section{Conclusion}
Predictions of 2-component model of the cosmic ray origin
and QGSJET high energy interaction model allow us to explain
the measured EAS size spectra in the knee region with  
accuracy better than 10\%. \\
\begin{figure}[t]
 \includegraphics[width=8.3cm,height=9cm]{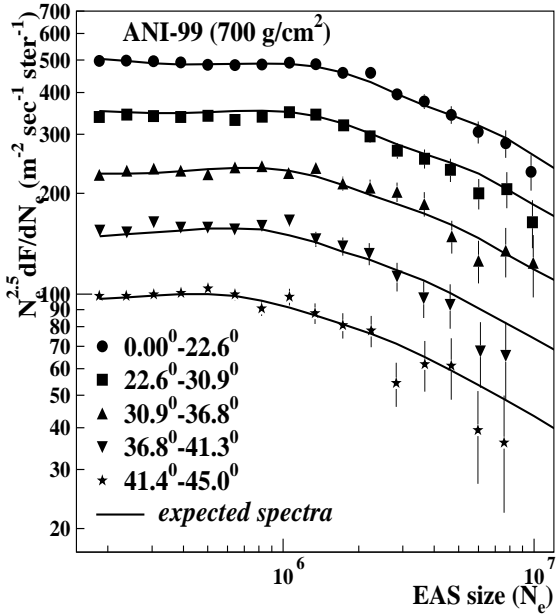} %
\caption{ANI EAS size spectra at different zenith angles 
\citep{ANI} (symbols). Lines correspond to expected EAS size
spectra
according to QGSJET interaction model and 2-component origin of 
primary cosmic rays.}
 \end{figure}
All model parameters of
2-component primary energy spectra, obtained from EAS data 
agree with theoretical predictions \citep{PB} in the frames
of standard errors with the exception of $\gamma_3$ parameters.
Obtained spectral slopes ($\gamma_3=3.28\pm0.07$) after
the knee of SW-component are significantly steeper than 
2-component model predictions (\citep{PB}, $3.07\pm0.1$)
and this result requires of further investigations.

\begin{acknowledgements}
The authors wish to thank many colleagues from experiments
ANI and KASCADE for providing EAS data. \\
The work has been partly supported by the research grant N 00-784
of the Armenian government,
NATO NIG-975436 and CLG-975959 grants and ISTC A216 grant.
\end{acknowledgements}

\end{document}